\documentclass[conference]{IEEEtran}
\IEEEoverridecommandlockouts
\usepackage{cite}
\usepackage{amsmath,amssymb,amsfonts}
\usepackage{algorithmic}
\usepackage{graphicx}
\usepackage{textcomp}
\usepackage{xcolor}
\begin{document}

\title{Age of Information in an URLLC-enabled Decode-and-Forward Wireless  Communication System
\thanks{This work is funded  by the Seed Funding program  of Sri Lanka Technological Campus, Sri Lanka grant no. RRSG/20/A7 and  by the Russian Foundation Basic Research grant № 19-37-90037.}
}
\author{\IEEEauthorblockN{Chathuranga M. Wijerathna Basnayaka}
\IEEEauthorblockA{\textit{Centre for Telecommunications Research} \\
\textit{Sri Lanka Technological Campus}\\
Padukka, SRI LANKA \\
chathurangab@sltc.ac.lk}
\and
\IEEEauthorblockN{Dushantha Nalin K. Jayakody}
\IEEEauthorblockA{\textit{School of Computer Science and Robotics} \\
\textit{National Research Tomsk Polytechnic University}\\
 Tomsk,RUSSIA\\
nalin.jayakody@ieee.org}
\and
\IEEEauthorblockN{Tharindu D. Ponnimbaduge Perera}
\IEEEauthorblockA{\textit{School of Computer Science and Robotics} \\
\textit{National Research Tomsk Polytechnic University}\\
Tomsk, RUSSIA \\
ponnimbaduge@tpu.ru}
\and
\IEEEauthorblockN{Moisés Vidal Ribeiro}
\IEEEauthorblockA{\textit{Department of Electrical Engineering} \\
\textit{Federal University of Juiz de Fora Campus Universitário}\\
Plataforma 5, BRAZIL \\
mribeiro@engenharia.ufjf.br}
}
\maketitle
\begin{abstract}
Age of Information (AoI) measures the freshness of data in mission critical Internet-of-Things (IoT) applications i.e., industrial internet, intelligent transportation systems etc. In this paper, a new system model is proposed to estimate the average AoI (AAoI) in an ultra-reliable low latency communication (URLLC) enabled wireless  communication system with decode-and-forward relay scheme  over the quasi-static Rayleigh block fading channels. Short packet communication scheme is used to meet both reliability and latency requirements of the  proposed wireless network. By resorting finite block length information theory, queuing theory and stochastic processes, a closed-form expression for AAoI is obtained. Finally, the impact of the system parameters, such as update generation rate, block length and block length allocation factor on the AAoI are  investigated. All results are validated by the numerical results.
\end{abstract}
\section{Introduction}
 The concept of age of information (AoI) as a data freshness metric was introduced in the early 90s by data scientists to evaluate data in terms of temporal consistency in real-time databases. The term named ``AoI'' was initially used by Kaul and Marco in their work on vehicular communication networks\cite{kaul2011minimizing}. With the rapid advancements of information technology, the idea of AoI was spread to other research domains such as wireless communication, since the dissemination of timely information became increasingly critical for modern wireless communication applications\cite{perera2020age}. On the other hand, wireless cooperative communication technologies have been vigorously investigated in the telecommunication domain due to its added advantages, i.e, improved reliability of transmissions and latency requirement of the network.\par Existing works in the literature related to AoI is concerned with single-hop wireless networks assuming each communication nodes is capable of transmitting data  with an infinite block length regime \cite{abdel2019optimized}.However, transmitting information involving an infinite-block length regime is impractical in the presence of URLLC-enabled wireless communication networks. Therefore, short-packet data transmission is used to meet both reliability and latency requirements of the wireless communication networks. Further, it is also noteworthy that in short-packet communication scenario, the Shannon-Hartley Capacity theorem is no longer applicable \cite{5452208}. In addition, integrating finite-block length regime into  wireless communication networks provide a precise framework to determine the relationship between the latency and the reliability between communication nodes. \par To the best of the authors knowledge, there is no existing work found that studies the AAoI performance metric of a wireless decode-and-forward (DF) relay scheme under finite block length regime. To fill this research gap, in this paper studies the AAoI in a DF relay-assisted wireless communication system considering a finite block length regime over a quasi-static Raleigh fading channel conditions. Consequently, AAoI in wireless networks is analyzed using queuing theory and stochastic processes, and then a closed-form expression for AAoI is derived. Finally, in order to estimate AAoI in the proposed system, a closed-form expression of the block error probability is incorporated into derived closed-form expression of AAoI.\par Our numerical simulation results proved that by selecting a small packet size when the update size is equal or less than $10$ bits always  help to maintain the URLLC with low AAoI in the proposed system. However, if the update size is greater than $10$ bits, optimal block length that achieved lowest AAoI is approximately equals to $20$ channel uses. It is also noteworthy that the lowest AAoI is achieved when the number of update generation rate at the source is equals to $22$ updates per second when the each update size is equals to $100$ bits. \par The remaining part of this paper is organized as follows. In Section II, the system model is presented  with the  derivations for closed-form expression of the block error probability for the proposed system. In the section III, AAoI of the proposed system is derived. In Section IV, numerical simulation results are provided to verify the mathematical derivations derived in Section II and III. Finally, paper concludes in Section V.
\section{System Model}
\begin{figure}[!ht]
\includegraphics[width=\linewidth]{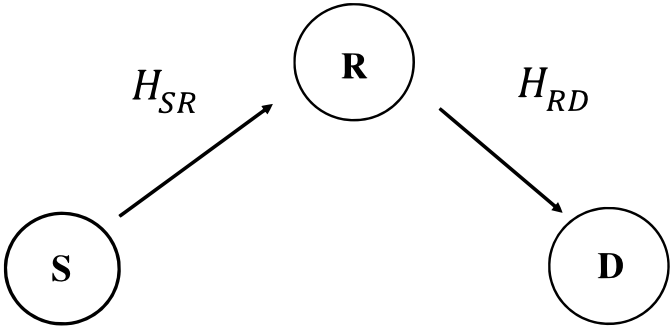}
\caption{System model for URLLC-enabled relay system}
\label{f1}      
\end{figure}
\par In this paper,  a URLLC-enable DF cooperative relaying system is proposed as illustrated in \figurename{\ref{f1}}, where the source $(S)$ sends its newly generated updates to the destination $(D)$ via aid of the relay $(R)$. In this proposed cooperative communication system, each transmission block is split into two separate time slots. The source sends data to the relay in the first time slot. Then, relay decodes and re-transmits the received data to the destination during the second time slot. It is assumed that there is no direct transmission between the source and the destination due to unfavourable line-of-sight (LoS) connection between them. The transmitted signal by the source, the received signal at the relay, the transmitted signal by the relay, and the received signal at the destination are denoted by $X_{1}$, $Y_{1}$, $X_{2}$ and $Y_{2}$, respectively. The received signal at each communication node can be written as
\begin{equation}
    Y_{1}=\sqrt{P_{S}} H_{SR}X_{1}+W_{SR},
\end{equation}
\begin{equation}
       Y_{2}=\sqrt{P_{R}}H_{RD}X_{2}+W_{RD},
\end{equation}
where $H_{ij}$ is the channel coefficient of the channel between node $i$ to node $j$ where  $i\in \left \{ S,R \right \}$ and $j\in \left \{ R,D \right \}$. The symbol $W_{ij}$ denotes the independent and identically distributed Addictive white Gaussian noise (AWGN) of the channel with zero mean and $\sigma^2$ variance. In this work, transmit power of the communication node $i$ is modelled as 
\begin{equation}
    P_{i} = \varphi_{i}P,
\end{equation}
where $P$ is the total transmission power constraint of the system and $ 0<\varphi_{i}\leq 1$ is the power allocation factor of each communication node. The distance between the source and the destination is considered  as $d$ and the distance between the source and the relay is represented by $d\tau$, where $0<\tau<1$. The small scale channel gain is denoted as $g_{ij}=|h_{ij}^2|$, where $h_{ij} \sim\mathcal{C}\mathcal{N}(0,1)$ denotes the Rayleigh fading channel coefficient. The probability density function (PDF) of the small scale channel gain $g_{ij}$ can be expressed as 
\begin{equation}
f_{g_{ij}}(z)=e^{-z}.
\end{equation}
The large scale channel gain $\alpha_{ij}$ between the communication nodes can be written as 
\begin{equation}
   -10log(\alpha_{\textit{i}\textit{j}} )= 20 log(d_{\textit{i}\textit{j}})+20log\bigg(\frac{4\pi f_{c}}{C}\bigg),
\end{equation}
where $f_{c}$ and $d_{ij}$ are the carrier frequency and distance between the communication nodes $i$ and $j$ respectively. The variable $C$ denotes the speed of light in the space. Considering small and large scale channel gains, the channel coefficient between the communication nodes can be written as
\begin{equation}
    H_{ij}=\sqrt{\alpha_{ij}g_{ij}}.
\end{equation}
The normalized received signal-to-noise ratio (SNR) $\gamma_{j}$ at the receiving node $j$ can be written as
 \begin{equation}
     \gamma_{\textit{j}} =\frac{ \alpha_{\textit{i}\textit{j}} g_{\textit{i}\textit{j}}P_{\textit{i}}}{\sigma_{ij}^{2}}.
 \end{equation}
The total block length $n$ allocated for the proposed system is is allocated for each transmission block according to the predefined allocation factor $\eta_{ij}$ between two communication nodes. The block length of each transmission block can be calculated as
\begin{equation}
   n_{ij}=\eta_{ij} n.
\end{equation} 
\subsection{ Block Error Probability in the Finite-Block Length Regime}
Considering DF relay protocol, the overall decoding error probability can be written as
\begin{equation}
    \varepsilon =\varepsilon _{R}+\varepsilon _{D}(1-\varepsilon _{R}),
    \label{frere}
\end{equation}
where $\varepsilon _{R}$ is the block error probability at the relay, $\varepsilon _{D}$ is the block error probability at the destination. Due to the static nature of the communication channels, it is assumed that the fading coefficients stay constant over the duration of each transmission block. Following Polyanskiy's results on short packet communication \cite{5452208} and assuming that the receiver has perfect channel state information, the expectation of the block error probability of a given block length can be written as
    \begin{equation}
        \varepsilon_{j}=\mathbb{E}\left [ Q\left ( \frac{n_{ij}C(\gamma_{j})-k}{\sqrt{n_{ij}V(\gamma_{j})}} \right ) \right ],
    \end{equation}
where  $\mathbb{E}\left [ . \right ]$ is the expectation operator, $Q(x)=\frac{1}{\sqrt{2\pi }}\int_{x}^{\infty }e^{-\frac{t_{2}}{2}}dt$ and $V(\gamma_{\textit{j}})$ is the  channel dispersion, which can be written $V(\gamma_{j} )=\frac{{\log _{2}}^{2}e}{2}(1-\frac{1}{(1+\gamma_{j} )^2})$. The variable $C(\gamma_{j})$ denotes the channel capacity of a complex AWGN channel and it is given by $C(\gamma_{j})=\log _{2}(1+\gamma_{j})$. The number of bits per block represents by $k$. Moreover, under the Rayleigh fading channel conditions, $\varepsilon_{j} $ can be formulated as
    \begin{equation}
      \varepsilon_{j} = \int_{0}^{\infty }f_{\gamma_{j}} (z)Q\left (\frac{n_{ij}C(\gamma_{j})-k}{\sqrt{n_{j}V(\gamma_{j})}} \right)dz,
      \label{e11}
    \end{equation}
where  $f_{\gamma_{j}}(z)$ denotes the PDF of the received SNR $\gamma_{j}$ at the $j^th$ communication node. Due to the complexity of the Q-function, it is difficult to get a closed-form expression for the overall decoding error probability. Thus, using the approximation technique given in \cite{makki2014finite}  and \cite{gu2017ultra}, \eqref{e11} can be approximated as
    \begin{equation}
        \varepsilon_{\textit{j}} \approx  \int_{0}^{\infty }f_{\gamma_{\textit{j} }} (z)\Theta_{\textit{j}} (z),
          \label{qfa1}
    \end{equation}
where $\Theta_{j}(z)$ denotes the linear approximation of $Q\left ( \frac{n_{ij}.C(\gamma_{j})-k}{\sqrt{n_{ij}.V(\gamma_{j})}} \right)$, this can be expressed as \cite{gu2017ultra}
    \begin{equation}
        \Theta_{j} (z)=\left \{ \begin{matrix}
1,&  \gamma_{j}\leq \phi_{j},  & \\ 
\frac{1}{2}-\beta_{j} \sqrt{n_{ij}}(\gamma_{j}-\psi_{j}), & \phi_{j} < \gamma_{j} <\delta_{j}, & \\ 
 0,&  \gamma_{j} \geq \delta_{j},& 
\end{matrix} \right.
\label{qfap}
    \end{equation}
where $\beta_{j} =\frac{1}{2\pi \sqrt{2^{\frac{2k}{n_{ij}}}-1}},\psi_{j}=2^{\frac{k}{n_{ij}}}-1,\phi_{j}=\psi_{j}-\frac{1}{2\beta_{j}\sqrt{n_{ij}}}$ and $\delta_{j}=\psi_{j}+\frac{1}{2.\beta_{j}\sqrt{n_{ij}}}$. Using  (\ref{qfa1}) and (\ref{qfap}), a closed-form expression for the block error probability can be derived as
\begin{equation}
    \varepsilon_{ij} \approx 1-\left (\frac{\beta_{j} \sqrt{n_{ij}}P_{i}\alpha_{ij}}{\sigma^2 } \right)\left( e^{-\frac{\phi_{i}\sigma^2 }{\alpha_{ij}P_{i}}}-e^{-\frac{\delta_{j}\sigma ^{2} }{\alpha_{ij}P_{i}}}\right).
    \label{rere2}
    \end{equation}
Finally, substituting \eqref{rere2} in  \eqref{frere}, a closed-form expression for the overall error probability can be obtained as 
\begin{equation}
\begin{aligned}
 \varepsilon \approx  &  1-\bigg[\left ( \frac{\beta _{R}\beta _{D}\sqrt{n\eta_{SR}\eta_{RD}}\times \varphi_{S} \varphi_{R} (P)^{2}\alpha _{SR}\alpha _{RD}}{\sigma ^{4}} \right ) \\ & \bigg(e^{-\frac{\phi_{R}\sigma ^{2}}{\alpha _{SR}\varphi_{S} P}}-e^{-\frac{\delta _{R}\sigma ^{2}}{\alpha _{SR}\varphi_{S} P}}\bigg)\bigg(e^{-\frac{\phi_{D}\sigma ^{2}}{\alpha _{RD}\varphi_{R}P }}-e^{-\frac{\delta _{D}\sigma ^{2}}{\alpha _{RD}\varphi_{R}P}}\bigg)\bigg].
\end{aligned}
\label{fere} 
\end{equation}
\section{ Average Age of Information of the proposed URLLC-enabled relaying scheme}
 If the generation time of the freshest update received at time stamp $t$ is $g(t)$, then AoI can be defined as a random process as 
\begin{equation}
   \Delta (t)=t-g(t).
\end{equation}
\begin{figure}[hb!]
\includegraphics[width=\linewidth]{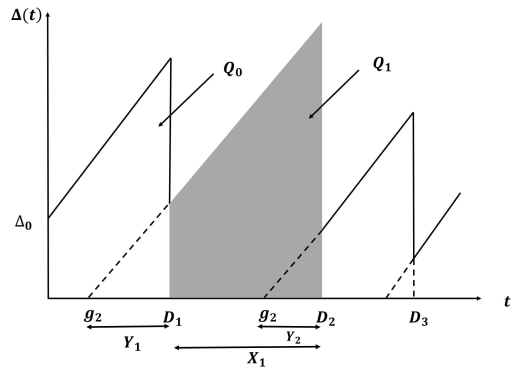}
\caption{ Evolution of Age of Information $(\Delta(t))$ with the time  }
\label{f2a}      
\end{figure}
As illustrated in the \figurename{ \ref{f2a}}, it is assumed that at $t=0$ the measurements of the AoI starts and the AoI at the destination is set to  $\Delta (0)=\Delta _{0}$. The information source generate updates at time stamps $\mathit{g_{1},g_{2},...,}$ and  the destination receive these updates at time stamps $\mathit{D_{1},D_{2},...,}$. As illustrated in \figurename{ \ref{f2a}}, data update $i$ is transmitted from the source at time stamp $t = g_{i}$ and it is successfully delivered to the destination at time stamp $D_{i}= g_{i}+Y_{i}$. Therefore, the AoI at the destination can be calculated as
 \begin{equation}
        \Delta (D_{i})=D_{i}-g_{i}.
        \end{equation}
AoI increases linearly until the next update is delivered to the destination. Similarly, AoI just before $i+1$ update is successfully delivered, can be written as
\begin{equation}
    \Delta  (D_{i+1}^{-})=Y_{i}+X_{i}. 
\end{equation}

Then, at the time stamp $D_{i+1}$, the AoI drops as follows
    \begin{equation}
        \Delta \left ( D_{i+1} \right )=Y_{i+1}. 
    \end{equation}
Considering the observation time interval as  $[0,T_{n}]$ for the proposed system, upper bound of the time interval can be given  as a function of $X_{i}$, which can be written as
    \begin{equation}
        T_{n}=\sum_{i=0}^{n-1}X_{i}.
        \label{tni}
    \end{equation}
For the considered time period, time average AoI can be computed using the area under $\Delta(t)$ and it can be treated as a  sequence of trapezoids  $Q_{i} $ when $0\leq i\leq n-1$ as shown in the \figurename{ \ref{f2a}}. Similarly, the time average age of the proposed system can be given as 
    \begin{equation}
      \Delta_{{T_{n}}}= \frac{1}{T_{n}} \int_{0}^{T_{n}} \Delta (t)d(t)=\frac{\sum_{i=0}^{n-1}Q _{i}}{T_{n}},
       \label{qi}
    \end{equation}
where
     \begin{equation*}
        Q_{i}=\int_{D_{i}}^{D_{i}+X_{i}}\Delta (t)d(t)=Y_{i}X_{i}+\frac{1}{2}X_{i}^2.
        \label{q}
    \end{equation*}
 Similar to the work presented in  \cite{kosta2017age}, the time average age ($\Delta_{{T_{n}}}$) tends to ensemble average age when  $n\rightarrow \infty $, i.e., which can be expressed as
 \begin{equation}
  \Delta_{a}=\lim_{n\rightarrow \infty }\Delta_{{T_{n}}}.
      \label{aaoie}  
       \end{equation}
Since $\mathbb{E}[Y_{i}]=\mathbb{E}[Y_{i+1}]$, $X_{i}$  and  update generation time difference have the same distribution. Thus, substituting  \eqref{tni} and \eqref{qi} to \eqref{aaoie}, the AAoI can be written as  
    \begin{equation}
        \textrm{AAoI}=\Delta_{a}=\frac{\frac{\mathbb{E}[X^{2}]}{2}+\mathbb{E}[XY]}{\mathbb{E}[X]}.
        \label{aifeq}
    \end{equation}
In \eqref{aifeq},  $\mathbb{E}[XY]$ is difficult to estimate without proper approximation since $Y$ and $X$ are dependent random variables. Thus, it is assumed that the communication node's data buffers are equipped with $M/G/1$ queue, where the source generate updates according to a Poisson process at the  rate $\lambda$ while the service time of the update is geometrically distributed. The system delay $(Y)$ can be calculated as a summation of the service time $(s)$  and waiting time $(w)$ of the queue, which can be written as
 \begin{equation}
     Y = s+w.
 \end{equation}
 Since $X$ and $s$ are i.i.d random variables,  AAoI is given by the \eqref{aifeq} can be re-formulate as 
\begin{equation}
  \Delta_{a} = \frac{\frac{1}{2}\mathbb{E}\left [ X^{2} \right ]}{\mathbb{E}\left [ X \right ]}+\mathbb{E}\left [ s \right ]+\frac{}{}\frac{\mathbb{E}\left [ wX \right ]}{\mathbb{E}\left [ X \right ]}.
  \label{afe3}
\end{equation}
\par Since $w$ and $X$ are dependent random variables, Pollaczek–Khinchine \cite{sac2018age,talak2018can} formula is used to further simplify \eqref{afe3}. Considering M/G/1 queue policy, the relationship between expectation of the variables $w$ and $s$  have been formulated as
\begin{equation}
    \mathbb{E}[w]=\frac{\mathbb{E}[s^{2}]}{2(\mathbb{E}\left [ X \right ]-\mathbb{E}[s])},
    \label{pkf}
\end{equation}
The correlation term $\mathbb{E}[wX]$ can be evaluated as \cite{inoue2017stationary,vlasiou2007non}
 \begin{equation}
     \begin{aligned}
       \mathbb{E}[wX]= & \frac{\mathbb{E}[X](\mathbb{E}[X]-\mathbb{E}[s])}{\mathbb{E}[e^{-\lambda s}]}+\frac{\mathbb{E}[s^2]\mathbb{E}[X]}{2(\mathbb{E}[X] -\mathbb{E}[s])} \\ &-\frac{\mathbb{E}[X^2]}{2},
     \end{aligned}
      \label{fae}
 \end{equation}
where $\lambda$ is the update generation rate, where $\mathbb{E}[X]=\frac{1}{\lambda }$ and $\mathbb{E}[X^2]=\frac{2}{\lambda^2}$. Then, using \eqref{fae} and \eqref{afe3}, AAoI can be written as
\begin{equation}
    \Delta_{a}=\mathbb{E}[s]+\frac{\mathbb{E}[s^2 ]}{2(\mathbb{E}[X]-\mathbb{E}[s])}+\frac{\mathbb{E}[X]-\mathbb{E}[s]}{\mathbb{E}[e^{-\lambda s}]}.
    \label{faoie}
\end{equation}
\par The variable $s$ in \eqref{faoie} depends on number of re-transmissions between the communication nodes. The total number of transmission $R$ needed for the reliable transmission of an update is geometrically distributed with transmission success rate and its probability mass function can be given by
\begin{equation}
    P_{R(m)}= (1-\varepsilon )\varepsilon ^{m-1}; m=1,2....
\end{equation}
where $\varepsilon_{ij}$ denotes the block error probability. The service time for an update, which is transmitted using one transmission block is given by
\begin{equation}
  s= (nT+\upsilon)R,
\end{equation}
 where  $n$ and $R$ denote block length and number of transmission needed for reliable communication, respectively, $\upsilon$ is the channel induced delay and $T$ is the symbol duration. Since all of these variables are constants and the service time is a  geometric random variable, thus, we have
\begin{equation}
    E(s)=\frac{nT+\upsilon}{1-\varepsilon},
    \label{aae1}
\end{equation}
 \begin{equation}
   E(s^{2})=\frac{(nT+\upsilon)^2(1+\varepsilon) }{(1-\varepsilon )^2},
   \label{aae2}
\end{equation}
\begin{equation}
  E(e^{-\lambda s})=\frac{(1-\varepsilon ) e^{-(nT+\upsilon)\lambda }}{1-\varepsilon e^{-(nT+\upsilon)\lambda}}.
  \label{aae3}
 \end{equation}
Finally, by using \eqref{fere}, \eqref{faoie}, \eqref{aae1}, \eqref{aae2} and \eqref{aae3} the AAoI of the proposed system can be derived as 
\begin{equation}
    \begin{aligned}
      {\Delta_{a}}=   \frac{(nT+\upsilon)}{1-\varepsilon} +\frac{\frac{(nT+\upsilon)^2{(1+\varepsilon) }}{(1-\varepsilon )^2}}{2(\frac{1}{\lambda}-\frac{(nT+\upsilon)}{1-\varepsilon})}+\frac{(\frac{1}{\lambda}-\frac{(nT+\upsilon)}{1-\varepsilon})}{\frac{(1-\varepsilon )e^{-(nT+\upsilon)\lambda }}{1-\varepsilon e^{-(nT+\upsilon)\lambda}}}.
    \end{aligned}
\end{equation}
\section{Numerical Simulation Results}
In this section, numerical simulation results are presented to verify the theoretical analysis presented in this paper. First, the effect of the update generation rate on the AAoI is investigated. Unless otherwise stated, the simulation parameters  are set as: $d= 1$ km, $P=23 $ dBm, $n=300$,  $\eta_{SR}=\eta_{RD}=0.5$, $\varphi_{S}=\varphi_{R} =0.5$, $\tau= 0.5$, $\sigma^{2}= -167$ dBm, $f_{c}= 6$ GHz, $C= 3\times10^8$ m/s, $k= 100$ and $T= 1\times10^{-4}$ s. It is assumed that the channel induced delay ($\upsilon$) is negligible compared to $nT$.
\begin{figure}[!htbp]
\includegraphics[width=\linewidth]{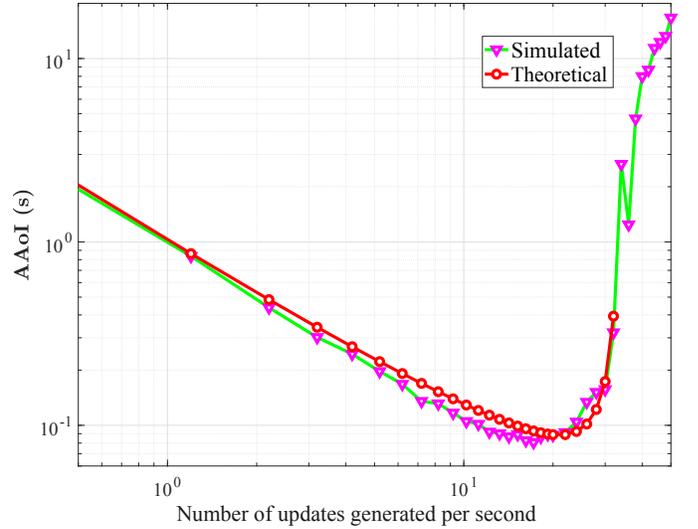}
\caption{AAoI as a function of update generate rate when $K=100$ bits.}
\label{sim1}      
\end{figure}
\par In \figurename{ \ref{sim1}}, AAoI is plotted as a function of update generation rate when $k=100$ bits. It can be observed from \figurename{ \ref{sim1}} that the network has a higher AAoI when the updates generation rate is either lower or greater than the optimal value of $22$ updates per second.This phenomena can be easily explained as follows. When the update generation rate is lower than the optimal value, it increases time difference between the reception of two consecutive updates causing an increase in AAoI. On the other hand, if the update generation rate significantly larger than the optimal value, it is impossible to stabilize the queue. Thus, packets are waiting for a longer period of time in the queue leading to a higher AAoI. Furthermore, it can be clearly seen from the figure that the simulation results are approximately similar to their analytical counterparts.
\begin{figure}[!htbp]
  \includegraphics[width=\linewidth ]{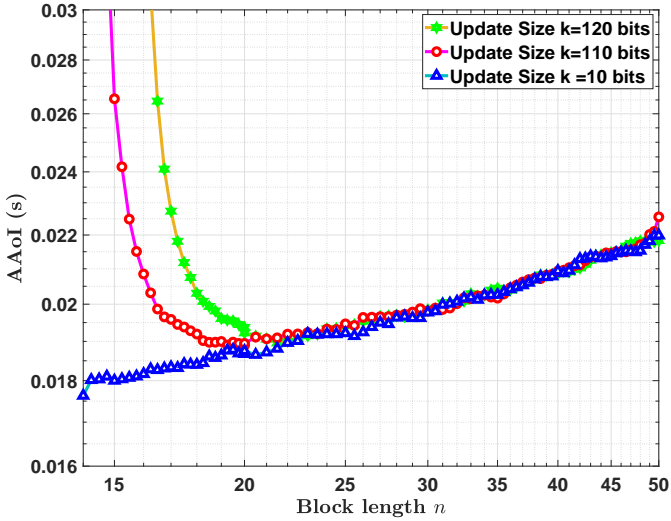}
\caption{ AAoI  as a function total block length $n$}
\label{sim2}      
\end{figure}
\par \figurename{ \ref{sim2}} illustrates the effect of the total block length $n$ on the AAoI performance of the proposed system. It is observed from the figure that when the update size $(k)$ is significantly small, the AAoI increase with the increases in block length. This is mainly due to the fact that when the  block length is increased, it causes increment in the service time for an update. On the other hand, an update consists of large number of bits has a higher AAoI when it is transmitted using block length less than $20$ channel uses due to the high block error probability. It is also noteworthy that when the update size is larger than $10$ bits, increase in block length towards its optimal value $20$ channel uses decreases the AAoI due to the decease in error probability. However, further increase in block length  after the optimal value has led to increase in AAoI due to the impact of the service time on AAoI is dominant compared to the decrease in block error probability. This result proved that transmitting small size update less than $10$ bits always help to maintain minimum AAoI for mission critical application. However, if the update size is greater than $10$ bits, lowest AAoI is achieved only when the block length is equals to $20$ channel uses. 
\begin{figure}[!htbp]
  \includegraphics[width=\linewidth ]{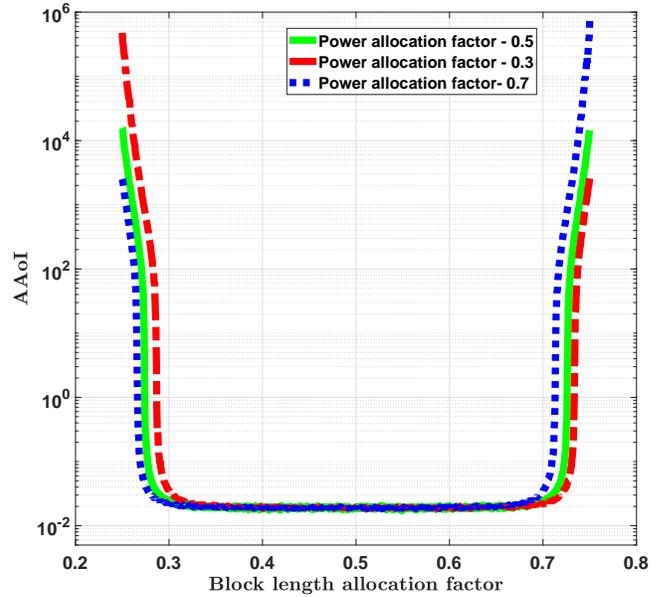}
\caption{ AAoI  as a function of block length allocation factor $\eta_{SR}$ for different power allocation factors at the $S$}
\label{sim3}      
\end{figure}
\par In \figurename{ \ref{sim3}}, AAoI is plotted as a function of the block length allocation factor of the $S$-$R$ link $(\eta_{SR})$ for different power allocation factors. It is shown that when the block length allocation factor increases, the AAoI decreases until it reaches the optimal value. This is mainly due to the fact that increase in block allocation factor decreases the block error probability at the $R$ compared to the block error probability at $D$. Thus, overall block error probability is reduced, which can be easily understood by referring to \eqref{frere}.When the block length allocation factor between  $0.3$ - $0.7$, the relay scheme experience lower AAoI, due to the moderate block error probabilities and stability of the queue. However, after the value $0.7$ the AAoI increases rapidly since the small block length allocation for a any channel increase the overall error probability. In addition, it is also shown that equal power allocation in communication nodes results in better AAoI regardless of the block length. 
\section{CONCLUSIONS}
This work developed  a new model to measure the AAoI in a DF wireless relay operating under ultra-reliable and low-latency constraints. The relationship between the block error probability and AAoI is formulated  within the finite
block length regime. In addition, the impact of update size on the AAoI is investigated and it has been shown that maintaining a small update size (short packet communication) minimizes the AAoI in wireless relay networks. In addition, the impact of the update generate rate, the block length allocation factor and the power allocation factor on the AAoI were also investigated.


\begin{thebibliography}{00}
\bibitem{kaul2011minimizing} S. Kaul, M. Gruteser, V. Rai, and J. Kenney, “Minimizing age
of information in vehicular networks,” in 2011 8th Annual IEEE Communications Society Conference on Sensor, Mesh and Ad HocCommunications and Networks, 2011, pp. 350–358.
\bibitem{perera2020age} T. D. P. Perera, D. N. K. Jayakody, I. Pitas, and S. Garg, “Age of
information in swipt-enabled wireless communication system for 5gb,” IEEE Wirel. Commun., vol. 27, no. 5, pp. 162–167, 2020.
\bibitem{abdel2019optimized} M. K. Abdel-Aziz, S. Samarakoon, C.-F. Liu, M. Bennis, and W. Saad, “Optimized age of information tail for ultra-reliable low-latency communications in vehicular networks,” IEEE Trans. Commun., vol. 68, no. 3, pp. 1911–1924, 2019.
\bibitem{5452208} Y. Polyanskiy, H. V. Poor, and S. Verdu, “Channel coding rate in the
finite blocklength regime,” IEEE Trans. Inf. Theory, vol. 56, no. 5, pp. 2307–2359, 2010.
\bibitem{makki2014finite} B. Makki, T. Svensson, and M. Zorzi, “Finite block-length analysis of the incremental redundancy harq,” IEEE Wirel. Commun. Le., vol. 3, no. 5, pp. 529–532, 2014.
\bibitem{gu2017ultra} Y. Gu, H. Chen, Y. Li, and B. Vucetic, “Ultra-reliable short-packet
communications: Half-duplex or full-duplex relaying?” IEEE Wirel. Commun., vol. 7, no. 3, pp. 348–351, 2017.
\bibitem{kosta2017age} A. Kosta, N. Pappas, and V. Angelakis, “Age of information: A new
concept, metric, and tool,” Foundations and Trends in Networking, vol. 12, no. 3, pp. 162–259, 2017.
\bibitem{sac2018age} H. Sac, T. Bacinoglu, E. Uysal-Biyikoglu, and G. Durisi, “Age-Optimal
Channel Coding Blocklength for an M/G/1 Queue with HARQ,” in 2018 IEEE 19th International Workshop on Signal Processing Advances in Wireless Communications (SPAWC), 2018, pp. 1–5.
\bibitem{talak2018can} R. Talak, S. Karaman, and E. Modiano, “Can determinacy minimize age
of information?” arXiv preprint arXiv:1810.04371, 2018.
\bibitem{inoue2017stationary} Y. Inoue, H. Masuyama, T. Takine, and T. Tanaka, “The stationary distribution of the age of information in FCFS single-server queues,” in 2017 IEEE International Symposium on Information Theory (ISIT), 2017, pp. 571–575.
\bibitem{vlasiou2007non} M. Vlasiou, “A non-increasing Lindley-type equation,” Queueing Systems, vol. 56, no. 1, pp. 41–52, 2007.
\end{thebibliography}
\end{document}